\documentclass[final,3p,times,onecolumn,number,review,reviewsort&compress]{elsarticle}
\usepackage{amssymb}
\usepackage{upgreek}
\usepackage{lineno}
\usepackage{hyperref}
\usepackage{makecell}
\usepackage{multirow}
\usepackage{caption}
\usepackage{subcaption}
\usepackage{graphicx}
\usepackage{array}
\usepackage{sectsty}

\journal{Astroparticle Physics}

\sectionfont{\fontsize{14}{20}\selectfont}
\subsectionfont{\fontsize{10}{18}\selectfont\itshape}

\begin{document}

\begin{frontmatter}

\title{Measurement of cosmic muon-induced neutron background with ISMRAN detector in a non-reactor environment}

\author[a,b]{R.~Dey}
\ead{neuphyroni@gmail.com}
\cortext[cor1]{Corresponding author}
\author[a]{P.~K.~Netrakanti}
\author[a]{D.~K.~Mishra}
\author[a]{S.~P.~Behera}
\author[a]{R.~Sehgal}
\author[a,b]{V.~Jha}
\author[b,c]{and L.~M.~Pant}

\address[a]{Nuclear Physics Division, Bhabha Atomic Research Centre, Trombay, Mumbai - 400085}
\address[b]{Homi Bhabha National Institute, Anushakti Nagar, Mumbai - 400094}
\address[c]{Technical Physics Division, Bhabha Atomic Research Centre, Trombay, Mumbai - 400085}

\begin{abstract}
The Indian Scintillator Matrix for Reactor Anti-Neutrinos (ISMRAN) is an above-ground, very short baseline reactor anti-neutrino (${\overline{\ensuremath{\nu}}}_{e}$) experiment, located inside the Dhruva research reactor facility, Mumbai, India. The primary goal of the ISMRAN experiment is the indirect detection of reactor ${\overline{\ensuremath{\nu}}}_{e}$ through an inverse beta decay (IBD) process, using a cluster of 90 optically segmented plastic scintillator detectors, weighing $\sim$1 ton. However, the most difficult to distinguish correlated background for the ISMRAN experiment is from fast neutrons, which cannot be actively rejected and as a consequence mimics the IBD process through proton recoil inside the detector's volume. In this work, we present the neutron capture time response and energy deposition of neutron capture signals generated by cosmic muons in the ISMRAN geometry, and we compare these experimental results with Geant4-based Monte Carlo (MC) simulations. The obtained mean capture time of fast neutrons is 74.46 $\pm$ 5.98 $\mathrm{\mu}$s and is comparable with the MC simulation results. The efficiency-corrected rate of muon-induced neutron background inside the ISMRAN geometry, due to the presence of a passive shielding structure of 10 cm lead followed by 10 cm borated polyethylene with a surface area of 600 $\mathrm{cm^{2}}$, deployed on top of the ISMRAN setup, is reported to be 1334 $\pm$ 64 (stat.) $\pm$ 70 (sys.) per day. This result shows good agreement with the expected background rate from MC simulations using Geant4. Additionally, we also estimate the muon-induced fast-neutron rate in the ISMRAN geometry for the actual shielding configuration of 9000 $\mathrm{cm^{2}}$ surface area to be 3335 $\pm$ 160 (stat.) $\pm$ 175 (sys.) neutrons $\mathrm{day^{-1}}$ through an extrapolation, after incorporating the model dependent acceptance correction factor from the Geant4 MC simulation. Finally, using these results, we evaluate the neutron production yield due to the composite shielding in the ISMRAN geometry, which is 2.81$\times$$\mathrm{10^{-5}}$ neutrons per $\mu$ per (g/$\mathrm{cm^{2}}$) at sea level. These results will be significant in the context of differentiating correlated background from true ${\overline{\ensuremath{\nu}}}_{e}$ events at the actual measurement site inside the reactor facility.


\end{abstract}


\end{frontmatter}


\section{Introduction}
\label{sec1}
Neutrinos have become one of the most crucial entities that have been investigated intensively for their fascinating and unique features in recent decades. Nuclear reactors are intense sources of electron anti-neutrinos (${\overline{\ensuremath{\nu}}}_{e}$) and have played a significant role in studying the fundamental properties of neutrinos, including the first observation of ${\overline{\ensuremath{\nu}}}_{e}$~\cite{Cowan}, precise measurements of the neutrino mixing parameters $\Delta m^2_{21}$ and $|\Delta m^2_{31}|$, as well as the observation of neutrino oscillations driven by the $\theta_{13}$ parameter~\cite{PDG} within the three flavour framework. Several very short baseline experiments, like STEREO and PROSPECT~\cite{prospect,stereo} are reporting exciting results in the search for sterile neutrinos and the precise measurements of the reactor ${\overline{\ensuremath{\nu}}}_{e}$ energy spectrum from the fission of the isotope $\mathrm{{}^{235}U}$.


Furthermore, an initial discrepancy of 6$\%$ in the measured global reactor ${\overline{\ensuremath{\nu}}}_{e}$ flux is reported when compared with the theoretical models ~\cite{Mueller,Huber,Mention,Vogel}, which is known as ``Reactor Anti-neutrino Anomaly" (RAA). Additionally, a deviation from the expected reactor ${\overline{\ensuremath{\nu}}}_{e}$ energy spectrum has been observed as an excess in the reconstructed ${\overline{\ensuremath{\nu}}}_{e}$ energy spectra within the 5$-$7 MeV range, with a local significance of up to $\sim$4$\sigma$. This feature is known as the “Shape Anomaly” or “5 MeV Bump”~\cite{DBay_bump,DChooz_bump}. The PROSPECT experiment compared their measured reactor ${\overline{\ensuremath{\nu}}}_{e}$ spectra with the Huber-Müller (HM) $\mathrm{{}^{235}U}$ model and disfavored both the no-$\mathrm{{}^{235}U}$ and the all-$\mathrm{{}^{235}U}$ hypotheses at 3.7$\sigma$ and 2.0$\sigma$, respectively, to explain the ``Shape Anomaly"~\cite{prospect2}. In the near future, results from the JUNO reactor ${\overline{\ensuremath{\nu}}}_{e}$ experiment can be used to disentangle the mass ordering of neutrinos with a baseline of 53 km~\cite{juno}. Very short baseline experiments can also provide remote monitoring of the reactor thermal power and evolution of the fuel compositions inside the reactor core during reactor burnup in a non-intrusive way~\cite{SONGS,NUCIFER}.  

An array of plastic scintillator bars (PSBs), referred to as Indian Scintillator Matrix for Reactor Anti-Neutrinos (ISMRAN), has been installed and commissioned at the Dhruva research reactor facility, in the Bhabha Atomic Research Centre (BARC) to address the aforementioned fields of interest. The Dhruva research reactor operates at a thermal power of 100 $\mathrm{MW_{th}}$~\cite{DHRUVA}. It uses natural uranium fuel ($\mathrm{{}^{238}U}$:99.3$\%$; $\mathrm{{}^{235}U}$:0.7$\%$) and provides a maximum thermal neutron flux of $1.8 \times 10^{14} \mathrm{n/cm^{2}/s}$. The Dhruva reactor is primarily used for producing radioisotopes for medical, industrial, and research applications. The ISMRAN is mainly designed to measure the yields and energy spectrum of ${\overline{\ensuremath{\nu}}}_{e}$, through the inverse beta decay (IBD) process near the reactor core. Measuring the ${\overline{\ensuremath{\nu}}}_{e}$ energy spectrum from $\mathrm{{}^{235}U}$ fission in ISMRAN will contribute to resolving the “Shape Anomaly”~\cite{DB5MeV,RENO5MeV,HUBER5MeV}. Real-time monitoring of the reactor thermal power and evolution of fuel compositions inside the reactor core can be demonstrated through the ${\overline{\ensuremath{\nu}}}_{e}$ rates as a function of reactor burnup~\cite{PROS,DayaBayFuel,IAEA,Oguri}.

The ISMRAN is equipped with 90 plastic scintillator bars (PSBs) arranged in a matrix, with each PSB wrapped in Gadolinium (Gd) sheets and the whole setup enclosed within a combined shielding configuration and is located 18 m away from the reactor core. The passive shielding structure is composed of 10 cm thick lead (Pb) and 10 cm thick borated polyethylene (BPE) referred to as the actual shielding (which is installed at the actual experimental site), with a surface area of 9000 $\mathrm{cm^{2}}$ and a boron concentration of 30$\%$ by weight. It is designed to mitigate reactor backgrounds from $\gamma$-rays, fast neutrons, and thermal neutrons inside the reactor facility. From MC simulations it is estimated that the shielding configuration has a rejection of up to $\sim90\%$ for fast neutron events within the energy range of 1 MeV to 3 MeV ~\cite{ISMRAN}. By respecting the weight limit of massive shielding structures on the floor of the reactor facility, the above mentioned shielding configuration is adopted for the ISMRAN experiment. The complete ISMRAN detector setup is mounted on a movable base structure, which will allow us to manoeuvre it easily and make measurements at different distances from the reactor core in the baseline range of 13 m to 18 m. The PANDA experiment investigated a similar study with an array of PSBs at a stand off distance of $\sim$25 m from the reactor core and has reported promising results~\cite{Oguri}. A prototype detector array called mini-ISMRAN, which is composed of 1/16th of the full-scale ISMRAN was installed and successfully recorded data inside the reactor facility in the year 2018. A total of 218 $\pm$ 50 (stat.) $\pm$ 37 (sys.) ${\overline{\ensuremath{\nu}}}_{e}$ candidate events were obtained after analyzing 128 days of reactor ON (RON) dataset with the mini-ISMRAN~\cite{miniISMRAN} detector setup.




\begin{equation}\label{eq:ibd}
\mathrm{ \overline \nuup_{e} + p \rightarrow e^{+} + n}.
\end{equation}

\begin{equation}\label{eq:ibd2}
\mathrm{ E_{prompt} \simeq E_{\overline \nuup_{e}} + (M_{p} - M_{n} - M_{e}) + 2M_{e} = E_{\overline \nuup_{e}} - 0.78~MeV}.
\end{equation}

\begin{figure}[h]
\begin{center}
\includegraphics[width=14cm,height=5.8cm]{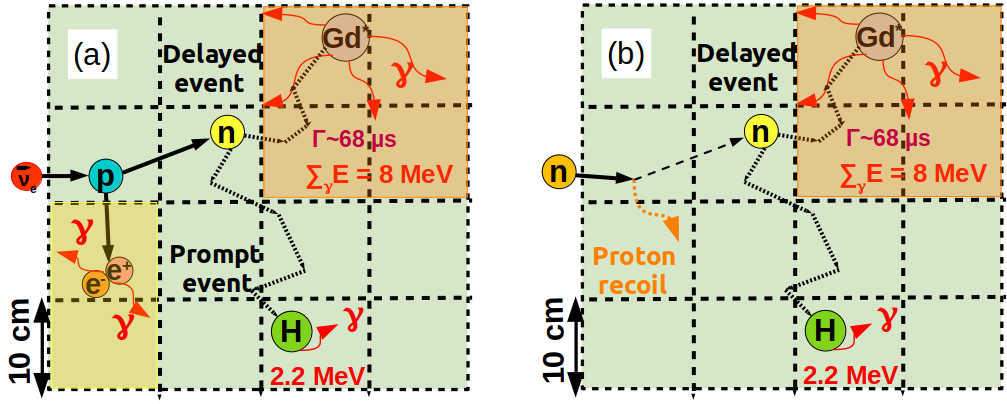}
\caption{ (a) Sketch for illustrating the IBD process. (b) Fast neutron event in a small section of the ISMRAN detector array.}
\label{fig1}
\end{center}
\end{figure}

The technique of detection of ${\overline{\ensuremath{\nu}}}_{e}$ events is through the IBD process, as explained in Eq.~\ref{eq:ibd}, where the positron deposits its energy in the PSBs and loses its energy. This process is followed by annihilation of the positron with the surrounding electrons in the material, resulting in the emission of two $\gamma$-rays of energy 0.511 MeV each. The positron energy deposition, along with contributions from the annihilation $\gamma$-rays, occurs within a very short timescale (within the resolution of our digitizer). Consequently, this is recorded as a composite signal and is referred to as the prompt signal. The prompt signal typically ranges in the energy scale from 0.8 MeV to 10 MeV, as detailed in Eq. \ref{eq:ibd2}. Eventually, a neutron, produced with an average energy of 10 keV from IBD, thermalizes in the PSB array and gets captured on either Gd sheets or hydrogen nuclei. The thermal neutron capture on Gd nuclei (n-Gd) results in the excitation of $\mathrm {Gd^{*}}$ nuclei and subsequent emission of a cascade of $\gamma$-rays through the de-excitation process. On the other hand, the neutron can also be captured by hydrogen nuclei (n-H), resulting in the emission of a mono-energetic $\gamma$-ray of energy 2.2 MeV. Since ISMRAN is an above-ground experiment, it is very challenging to discriminate the neutron capture on hydrogen nuclei signal from the high rate of natural $\gamma$-rays as well as cosmic muon-induced $\gamma$-ray background, which are discussed in detail in section~\ref{sec2.3}. Hence, we mainly focus on the neutron capture on Gd nuclei and call this the delayed signal. However, fast neutrons constitute a dominant background for the ISMRAN experiment, as they can easily mimic the IBD signal. Figure~\ref{fig1} (a) and Fig~\ref{fig1} (b) show the schematic view of part of the ISMRAN detector with an IBD event signature and a fast neutron event signature, respectively. Our simulations show the mean characteristic time ($\tau$) between the prompt and delayed event is $\sim$68 $\mu$s~\cite{ISMRAN} and is the same in both IBD and fast neutron events in the ISMRAN detector array. The mean characteristic time of IBD events is adopted from the Geant4-based MC simulation in the ISMRAN geometry~\cite{ISMRAN}. The overlapping energy range in the prompt events for both cases makes it even more difficult to discriminate the IBD signal from fast neutron and $\gamma$ background, which is discussed in section~\ref{sec5.2}.

The ISMRAN experiences correlated fast neutron backgrounds predominately produced by cosmic rays due to the interaction with high-density shielding material (Pb), surrounding the ISMRAN geometry~\cite{paddle1,muon_neutron}. High-energy primary particles coming from space collide with atoms in the atmosphere and generate high-energy nucleons and elementary particles. After interactions in the atmosphere and the decay of secondary particles, fast neutrons are created. Being an above-ground experiment, the ISMRAN detector setup can detect those fast neutrons that penetrate the shielding material. Fast neutrons can produce a prompt-like event signature through proton recoil energy deposition, followed by thermalization and eventually capture by the neutron capture agent present in our scintillator volume, providing a delayed-like event signature~\cite{paddle2}. On account of the scintillation quenching effect in the detector, those recoils (light output) are measured in the electron-equivalent energy range of the IBD process~\cite{D_D_Roni}. Cosmic muons, on the other hand, are produced through interactions of cosmic rays with particles in the atmosphere. Those high-energy muons can create fast neutrons through spallation reaction with high-atomic-number (Z) shielding material in the vicinity of the detector and can also mimic IBD signals~\cite{muon_neutron1}. Those kinds of backgrounds are known as cosmogenic background. In addition, the proximity of the ISMRAN detector setup to the reactor core comes at the cost of an additional reactor-induced fast neutron and $\gamma$-ray background, called reactogenic background. 


A small muon telescope system (sample shielding) was installed on top of the ISMRAN geometry in the detector laboratory to estimate the cosmic muon-induced fast neutrons in the shielding material. The sample shielding configuration used in this test setup is exactly the same as that used in the full-scale ISMRAN experiment inside the reactor facility (only the surface areas of shielding structures are different). Initially, these measurements were proposed to be carried out with the actual shielding of the ISMRAN detectors, but due to the weight issues and assembly complications in the detector laboratory, the idea was pursued with a smaller sample shielding test setup.


In the present work, we characterize the muon-induced neutron backgrounds in the ISMRAN geometry, due to the heavy shielding involved. The measurement of these secondary neutrons at specific places is quite difficult, and hence a sample shielding structure was devised to demonstrate the capability of such a system to estimate muon-induced neutron rates in the detector geometry. These studies have been performed at the detector laboratory, which is far away from the reactor facility, the so-called non-reactor environment. In addition, we also compared these measurements to the Geant4-based\cite{Geant4} (version 11.1.2) Monte-Carlo (MC) simulation to benchmark these background estimations. We have also evaluated the production yield of muon-induced neutron backgrounds in the ISMRAN geometry. 



\section{The ISMRAN Experiment}
\subsection{Experimental Setup and Data Acquisition System}
\label{sec2.1}
\begin{figure}[h]
\begin{center}
\includegraphics[width=16.4cm,height=7.4cm]{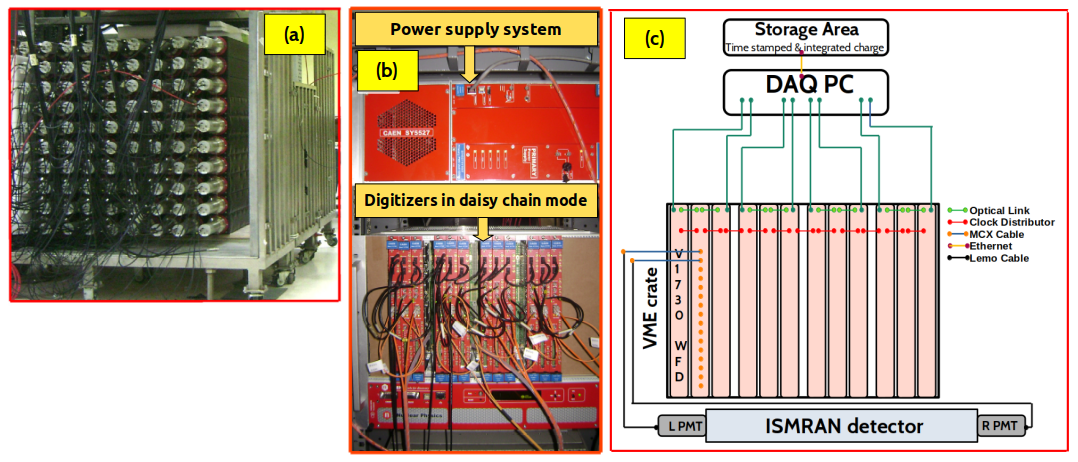}
\caption{(a) Full-scale ISMRAN detector setup consisting of 90 PSBs. (b) DAQ system for ISMRAN experiment. (c) Schematic view of the DAQ system.}
\label{fig2}
\end{center}
\end{figure}

The ISMRAN detector setup, as shown in Fig.~\ref{fig2} (a), is composed of an array of 90 Plastic Scintillator Bars, weighing $\sim$1 ton, arranged in the form of a matrix of 9$\times$10. An aluminized mylar sheet coated with gadolinium oxide ($\mathrm{Gd_{2}O_{3}}$) is wrapped around each PSB. The area density of the $\mathrm{Gd_{2}O_{3}}$ on these sheets is 4.8 $\mathrm{mg/cm^2}$. Each module has 10 kg of plastic scintillator (EJ-200)~\cite{eljen} and is 100 cm long with a cross-section of 10 $\times$ 10 $\mathrm{cm^2}$. At both ends of each PSB, 3-inch diameter PMT are coupled via optical grease for signal readout. The majority of PMTs, used in 70 PSBs, are the low-noise ETL 9305 series with a 10-stage dynode structure, while the remaining 20 are the high-gain ETL 9821 series with a 12-stage dynode structure~\cite{PMT1,PMT2}. The PSBs with high-gain PMTs are on the outermost top and side layers of the ISMRAN for the tagging of cosmic muons entering the detector setup. Fast signal processing and event recording are essential pre-requisite to avoid losing signal events in the presence of high rates of $\gamma$-rays and neutron backgrounds inside the Dhruva reactor facility around our experimental site. Therefore, the ISMRAN data acquisition system (DAQ) system uses nearly deadtime-less CAEN V1730 VME based waveform digitizers as the primary pulse processing component~\cite{digitizer}. It is a 16 channel and 14 bit ADC resolution waveform digitizer with a sampling rate of 500 million samples per second (MS/s). The nominal readout rate of our PSB is $\sim$150 events $\mathrm s^{-1}$. With rates as high as 4~$\times$ $\mathrm{10^{3}}$ events $\mathrm{s^{-1}}$ in each bar, tested with a $\mathrm{{}^{22}Na}$ radioactive source, there is practically no loss of events due to pileup or data transfer from the digitizer to the DAQ PC. To collect data independently from all the 90 PSBs of ISMRAN with practically zero dead time, 12 digitizers have been used in a time synchronized mode, as shown in Fig~\ref{fig2} (b). The pulse discrimination, trigger generation, threshold selection, charge integration in units of ADC, timestamp for each event using constant fraction discrimination (CFD) and coincidence of the PMT signals from each PSB are performed by a Digital Pulse Processing (DPP) algorithm using on-board FPGAs of the digitizers. The anode signals from the PMTs at both ends of a PSB are required to have a timing coincidence of 36 ns to be recorded as a coincidence event. The digitizer output is fed to the DAQ PC using optical fiber cables (which can transmit data at 80 MB/s) and Chainable Optical NETwork (CONET) cards, as illustrated in Fig~\ref{fig2} (c). The timestamped data from each PSB is then further analyzed offline using energy deposition, timestamp and position information to build an event~\cite{miniISMRAN}.

\subsection{Energy, Time and Position Responses of ISMRAN Detector}
\label{sec2.2}
Gain matching of PMTs has been carried out to achieve a uniform energy response among the 90 PSBs independently~\cite{roni_ISMRAN} and the gains are matched within 3$\%$ for PSBs. We also measured time and position resolutions, and the obtained timing resolution of $\sim$4 ns leads to a parameterized position resolution of $\sim$20 cm for ISMRAN detectors~\cite{ISMRAN}. However, the timing resolution of the PSBs improves for the reconstructed energy deposition ($\mathrm{E_{bar}}$) of high-energy cosmic muon events, as demonstrated in reference ~\cite{RAMAN}. Timing and position informations are useful in the context of filtering out signal events from accidental background events arising from the natural radioactivity. Reconstruction of prompt and delayed events in our analysis is performed based on their timing information and energy deposition in the ISMRAN detector array. Therefore, a good knowledge of the energy resolution ($14\%$ at 1 MeV with 2$\%$ systematic uncertainty), energy non-linearity ($4.5\%$ at 0.5 MeV with 3$\%$ systematic uncertainty), and timing resolution (3 ns with 2$\%$ systematic uncertainty) of our detector and PMT is required to reconstruct the IBD signal and backgrounds in the ISMRAN geometry~\cite{roni_ISMRAN}.

The measured average event rates for natural $\gamma$-rays (1 $<$ $\mathrm{E_{bar}}$ (MeV) $<$ 3) and cosmic muon backgrounds (15 $<$ $\mathrm{E_{bar}}$ (MeV) $<$ 50) in a PSB at the top surface of the ISMRAN geometry are $\sim$250 events $\mathrm s^{-1}$ and $\sim$10 events $\mathrm s^{-1}$, respectively. The rates are uniform over the period of time and almost consistent among all the 90 PSBs. Eventually, we evaluated the integrated cosmic muon flux (muons passing through a flat horizontal surface) from the muon event rate calculation, which is $ 0.95$ $\mathrm{cm^{-2}}\mathrm{min^{-1}}$, and it agrees well with the previously reported results from the ISMRAN experiment~\cite{shiba}.

\subsection{On-site $\gamma$-ray and Neutron Backgrounds}
\label{sec2.3}
\begin{figure}[h]
\begin{center}
\includegraphics[width=13.0cm,height=5.8cm]{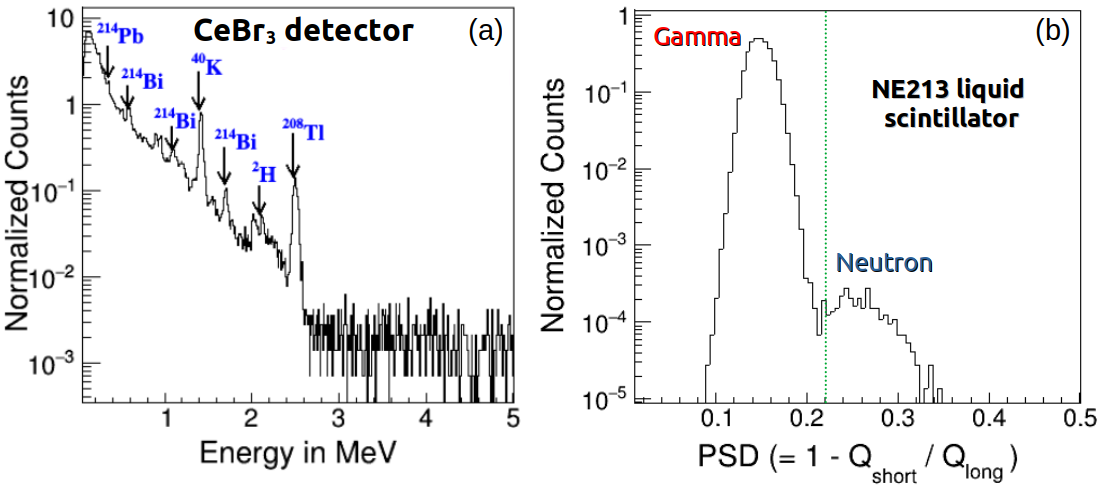}
\caption{The measured spectra of naturally occurring $\gamma$-rays and cosmogenic neutrons are shown in (a) and (b), respectively. Integrated charge in $\mathrm{T_{short}}$ denoted by $\mathrm{Q_{short}}$ and integrated charge in $\mathrm{T_{long}}$ denoted by $\mathrm{Q_{long}}$.}
\label{fig3}
\end{center}
\end{figure}

Additionally, we measured the energy spectrum of natural $\gamma$-rays, the integrated yield of the cosmogenic neutron background and neutron from concrete walls of the building via ($\alpha$,n) reaction using a 2$''$ diameter $\mathrm{CeBr_{3}}$ detector and a 5$''$ diameter by 2$''$ height NE213 liquid scintillator (LS) detector, which are shown in Fig~\ref{fig3} (a) and (b), respectively. Pulse shape discrimination (PSD) technique in LS involves using the integrated charge in long gate ($\mathrm{T_{long}}$) of 90 ns and short gate ($\mathrm{T_{short}}$) of 26 ns to identify fast neutron and $\gamma$-rays separately above 1 MeV. Figure.~\ref{fig3} (a) shows the $\gamma$-ray energy deposition spectrum in the $\mathrm{CeBr_{3}}$ detector. The prominent peaks in the measured spectrum are highlighted. The dominant $\gamma$-ray background comes from $\mathrm{{}^{40}K}$ and $\mathrm{{}^{208}Tl}$. Above 3 MeV, the natural $\gamma$-ray background remains almost constant as a function of energy. A high-energy tail can arise from the $\gamma$-rays, produced in the neutron capture process. The neutrons can arise from cosmic rays or spontaneous fission, or ($\alpha$, n) reactions. 

Figure~\ref{fig3} (b) shows the projection of PSD distribution for natural $\gamma$-rays and cosmogenic neutron background events recorded in LS detector of surface area 330 $\mathrm{cm^{2}}$. The centroid of natural $\gamma$-rays background distribution is at 0.15 and fast neutrons range from 0.22 to 0.35 in PSD distribution, peaking at around 0.28. From this study we measured the rate of cosmogenic neutron events at sea level and obtained a fast neutron rate $\sim$0.78 neutrons $\mathrm{day^{-1}}$.~$\mathrm{cm^{-1}}$~\cite{cosmic_neutron}. The average detection efficiency of fast neutron in LS is 34\%~\cite{LS_eff}. These measured results are essential prerequisites for estimating the accidental $\gamma$-ray and cosmogenic neutron background rates in a non-reactor environment.

\section{Expected Backgrounds for the ISMRAN Experiment}
\label{sec3}
\begin{figure}[h]
\begin{center}
\includegraphics[width=16cm,height=6.4cm]{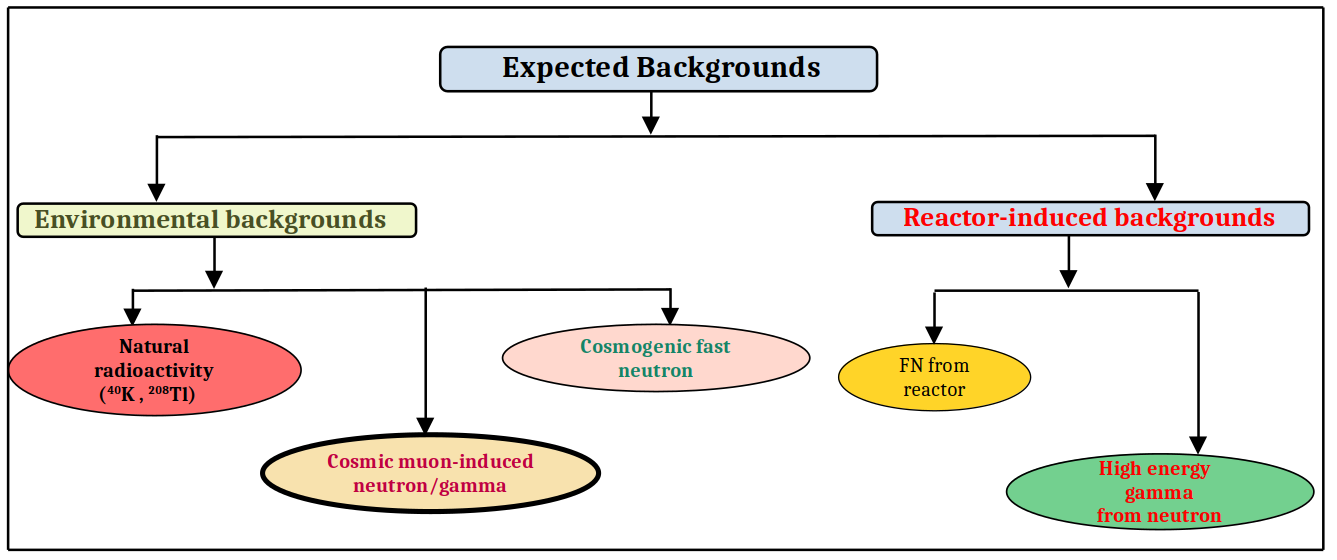}
\caption{ Schematic illustration of environmental and reactor-induced backgrounds for ISMRAN experiment.}
\label{fig4}
\end{center}
\end{figure}

The very small interaction cross sections of ${\overline{\ensuremath{\nu}}}_{e}$ demand a very precise knowledge of expected backgrounds for the search of ${\overline{\ensuremath{\nu}}}_{e}$. There are mainly two types of background for the ISMRAN experiment (a)  Environmental background and (b) Reactor-induced background, which are illustrated in Fig.~\ref{fig4}. The reactor-induced background is dominated by fast neutrons and high-energy $\gamma$-rays inside the reactor facility. The dominant sources of environmental backgrounds are the presence of the interfering cosmic muon-induced and natural $\gamma$-rays~\cite{paddle3}. The fast neutron correlated background is the most crucial background among all others, in the context of the purity of ${\overline{\ensuremath{\nu}}}_{e}$ candidate events~\cite{paddle4}.

\subsection{Different Sources for Fast Neutron Backgrounds}
\label{3.1}
Fast neutrons are mainly generated by the primary and secondary fission processes inside the reactor core and can escape from nearby experimental setups that utilize reactor beam ports. Furthermore, neutrons can also be emitted by ($\alpha$,n) reactions with the concrete walls of the reactor building and also via spontaneous fission. Alpha-emitting daughters from the trace isotopes of $\mathrm{{}^{238}U}$, $\mathrm{{}^{235}U}$ and $\mathrm{{}^{232}Th}$ inside the Dhruva reactor facility can produce neutrons through the ($\alpha$,n) reaction. 

Fast neutrons can also be generated when a low-energy negative muon is captured by a nucleus inside the detector due to Coulomb interaction, forming a ``muonic atom" bound state, resulting in the production of a neutrino and a neutron ( $\mathrm{\mu^{-}}$ + p $\mathrm{\rightarrow}$ n + ${\ensuremath{\nu}}_{\mu}$). Additionally, a high-energy muon can collide with a nucleus in a high-Z material and produces secondary particles. These high-energy muon can transfer some of their energy to the nucleus, causing it to break up into smaller fragments through a process known as muon-nucleon quasi-elastic scattering process. This spallation interaction can generate a variety of secondary particles, including protons, neutrons, pions, and other mesons. Moreover, electromagnetic showers generated by cosmic muons can also produce fast neutrons through various mechanisms. These high-energy photons in the shower can interact with nuclei, leading to the emission of neutrons. This process is known as photo-nuclear interaction. Cosmogenic nuclei such as $\mathrm{{}^{9}Li}$ and $\mathrm{{}^{8}He}$ contribute significantly to the irreducible backgrounds in reactor ${\overline{\ensuremath{\nu}}}_{e}$ experiments. Both isotopes undergo $\beta$-decay, which can excite their daughter nuclei $\mathrm{{}^{9}Be}$ and $\mathrm{{}^{8}Li}$, respectively. These excited states can release a neutron as they relax ($\mathrm{{}^{9}Li}$ $\rightarrow$ $\mathrm{{}^{9}Be}$ + $\mathrm{e^{-}}$ + ${\overline{\ensuremath{\nu}}}_{e}$ ; $\mathrm{{}^{9}Be}$ $\rightarrow$ $\mathrm{{}^{8}Be}$ + n and $\mathrm{{}^{8}He}$ $\rightarrow$ $\mathrm{{}^{8}Li}$ + $\mathrm{e^{-}}$ + ${\overline{\ensuremath{\nu}}}_{e}$ ; $\mathrm{{}^{8}Li}$ $\rightarrow$ $\mathrm{{}^{7}Li}$ + n or $\mathrm{{}^{8}Li}$ $\rightarrow$ $\mathrm{{}^{6}Li}$ + 2n), and the coincidence between the emitted $\beta$ particle and neutron can mimic the expected IBD signal. The Q-values for $\mathrm{{}^{9}Li}$ and $\mathrm{{}^{8}He}$ are 13.61 MeV and 10.7 MeV, respectively, with half-lives of 0.178 s and 0.119 s~\cite{cosmo_Li,cosmo_He}. They are challenging to reject using an anti-coincidence muon veto selection cut. The correlated backgrounds are also generated from $\mathrm{\mu^{-}}$ capture on $\mathrm{{}^{12}C}$, leading to fast neutron generation~\cite{karmen}.



 
 
 

\section{Detection Strategy of Cosmic Muon-induced Neutrons inside the ISMRAN geometry}
\label{4}
High-energy cosmic muons often produce spallation products, of which our interest lies in fast neutrons, contributing correlated backgrounds for the ISMRAN experiment. These fast neutrons are produced due to the muon-nucleus interaction of high-energy muons with the concrete wall of the reactor facility, surrounding high-Z material of the experimental site, or high-Z passive shielding material like Pb and also inside the detector volume~\cite{spa_neut}. The rate of this neutron background varies with the energy of the incoming cosmic muon. 

\begin{figure}[h]
\begin{center}
\includegraphics[width=15.4cm,height=5.6cm]{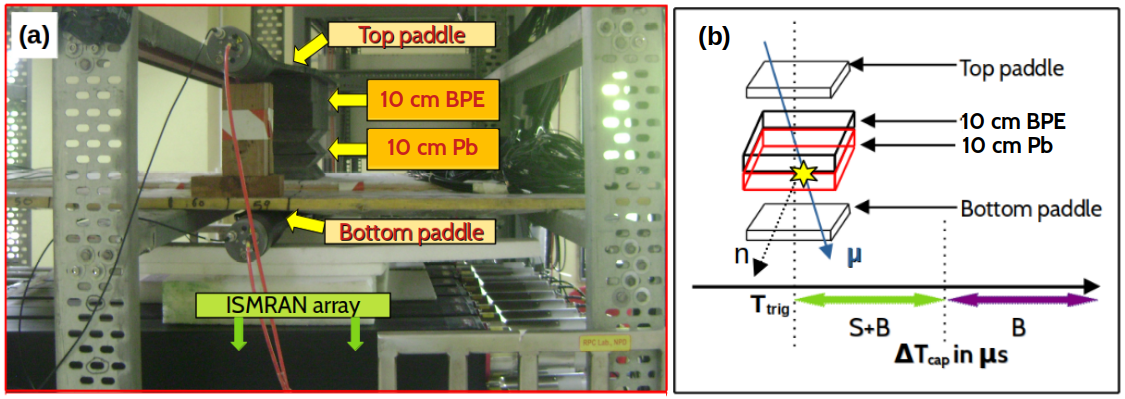}
\caption{(a) Experimental setup of the sample shielding structure with two plastic scintillator paddles from the muon telescope positioned on top of the ISMRAN geometry in a non-reactor environment. (b) Sketch for illustrating the muon-induced neutron detection strategy adopted for the ISMRAN experiment.}
\label{fig5}
\end{center}
\end{figure}


To evaluate the rate of muon-induced neutrons inside the ISMRAN detector setup due to the sample shielding material, a muon trigger system (muon telescope) is installed on top of the ISMRAN geometry, as shown in Fig~\ref{fig5} (a). The muon telescope system consists of two plastic scintillator paddles of dimension $\mathrm{20\times 30\times 2}$ $\mathrm{cm^{3}}$ and in between these two paddles, sample passive shielding are deployed. The surface area of 600 $\mathrm{cm^{2}}$ (20$\times$30 $\mathrm{cm^{2}}$), covered by the paddles, is used for the triggering of cosmic muon events. Fast neutrons are produced due to the interaction of high-energy cosmic muons with the sample shielding structure through the muon spallation process, while passing through these plastic scintillator paddles. This small muon telescope system was deployed on top of the ISMRAN geometry, since it was not possible to keep the large area of shielding structure due to the space and weight handling constraints of the setup. A schematic diagram of the detection technique of cosmic muon-induced neutrons is illustrated in Fig~\ref{fig5} (b) and described as follows:
 
 (1) Cosmic muons passing through the high-Z sample shielding material are triggered with plastic scintillator paddles.\par
 
 (2) These cosmic muons occasionally interact with the high-Z shielding material (Pb) producing fast neutrons of energy ranges from MeV to GeV, as by products. \par
 
 (3) Muon-induced neutron emerging from the high-Z shielding volume, entering into the ISMRAN detector setup and eventually captured on hydrogen nuclei providing 2.2 MeV mono-energetic $\gamma$-ray or on Gd nuclei, providing a cascade of $\gamma$-rays of sum energy 7.9 MeV after thermalization inside the ISMRAN detector array. \par
 
 (4) A cascade of $\gamma$-ray signal from neutron capture on Gd nuclei recorded within a pre-defined time window after the passage of a muon through the paddles. \par
 
 (5) The background is measured between the end of the correlated signal window and the passage of the next muon, which is explained schematically in Fig~\ref{fig5} (b).
 
\section{Geant4-based MC Simulation for Studying Muon-induced Neutron Background}
\label{5}


\subsection{Description of Physics List and Event Generator}
\label{5.1}
\begin{figure}[h]
\begin{center}
\includegraphics[width=16.4cm,height=5.2cm]{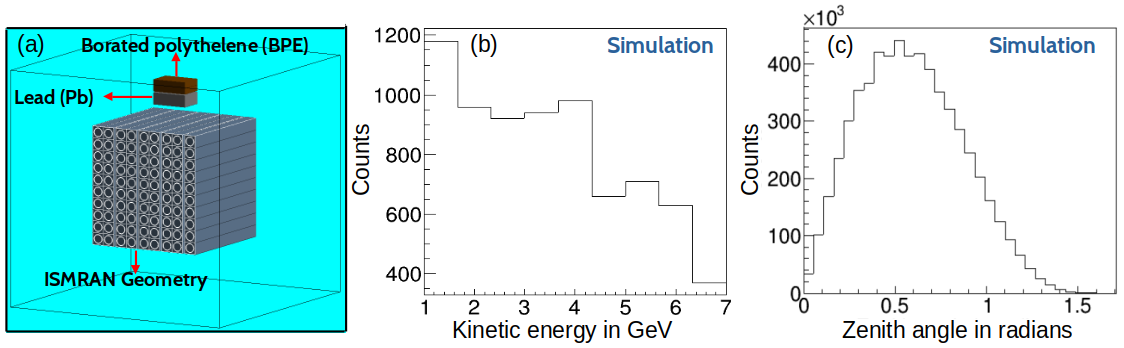}
\caption{ (a) Geant4-based detector simulation of the ISMRAN geometry, where the sample shielding structure, consisting of Pb and BPE, is used to study cosmic muon-induced neutron backgrounds. The kinetic energy and zenith angle distributions of cosmic muons generated by CRY on top of the sample shielding structure are shown in (b) and (c), respectively.}
\label{fig6}
\end{center}
\end{figure}

A Geant4-based simulation framework has been developed to investigate the effect of cosmic muon-induced backgrounds in the ISMRAN detector array (G4ISMRAN)~\cite{roni_ISMRAN,miniISMRAN}. We simulated the ISMRAN geometry of 90 PSBs (9$\times$10 matrix) with 10 cm thick Pb and BPE (surface area 20$\times$30 $\mathrm{cm^{2}}$) on top of the detector array using the Geant4 toolkit (version 11.1.2) which allowed us to mimic the real experimental setup, which is shown in Fig.~\ref{fig6} (a). Geant4 also offers a variety of physics lists i.e. sets of physics interactions which users can include as per their requirements in simulation. For G4ISMRAN, the primary physics requirements include the standard electromagnetic and radioactive decay physics processes for the response of $\gamma$-rays, electrons and positrons from radioactive sources. The high precision QGSP BIC HP physics process is used for high-energy hadron-nucleus and hadron-nucleon interactions in materials in the Geant4 MC simulation~\cite{Geant4}.


To achieve precise description of cascades of $\gamma$-ray emission from the thermal neutron capture on Gd nuclei in our MC simulation studies, we have implemented the DICEBOX software package which accurately models the emission of cascades of $\gamma $-rays from the region of high energy level density in an excited nucleus, based on the different models of energy level density and photon strength functions~\cite{Dicebox,Anigd}. High-energy particles such as cosmic muons and cosmic muon-induced neutrons also need to be simulated. Photo-Nuclear interaction (G4PhotoNuclearProcess), muon capture process (G4MuonMinusCapture), muon nucleus interaction (G4MuonNuclearProcess), and muon decay physics processes are explicitly included in the user-defined physics list to simulate muon-induced backgrounds and their interactions in the expected energy range in the ISMRAN geometry. 


Incoming high-energy cosmic muons can be generated with different cosmic muon generators, which are available as open-source codes to be integrated to particle transport code like Geant4. In the present work, we generated cosmic muons at sea level using the cosmic-ray air shower (CRAS) generator, which includes the cosmic ray shower library (CRY), linked with Geant4 as a particle transporter~\cite{cry}. CRY follows Gaisser’s ~\cite{Gaisser1,Gaisser2} parametrization with modification at lower energies and higher zenith angles. We generated randomly 55 million cosmic muon events using the CRY software on top of the ISMRAN geometry as well as sample shielding of Pb and BPE, passing into the Geant4 detector geometry, which is equivalent to 4.31 days of cosmic muon events (similar size to the amount of data taken).




\begin{figure}[h]
\begin{center}
\includegraphics[width=16.6cm,height=4.8cm]{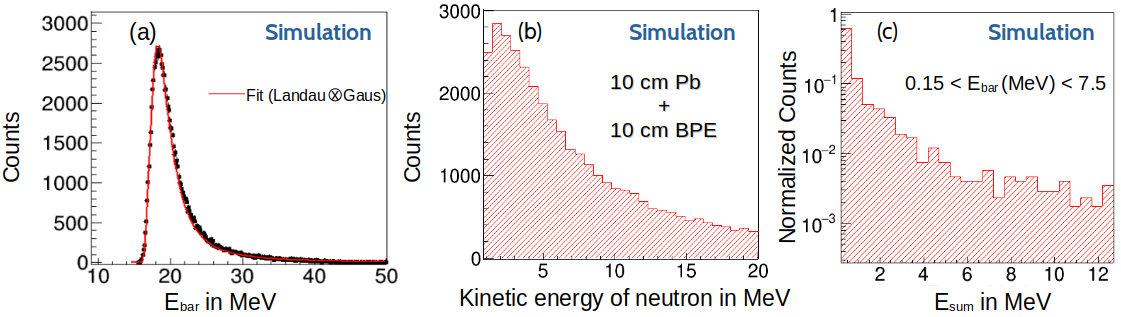}
\caption{(a) The energy deposition spectrum of cosmic muons passing through a single PSB. (b) The kinetic energy distribution of muon-induced neutrons. (c) Proton recoil energy deposition spectrum resulting from muon-induced neutrons within the detector array.}
\label{fig7}
\end{center}
\end{figure}

\subsection{Simulation of Cosmic Muon and Muon-induced Neutrons} 
\label{sec5.2}
CRY generates realistic momentum and angular distributions of cosmic muon at various altitudes. Figure~\ref{fig6} (b) and (c) show the kinetic energy and zenith angle distributions of cosmic muons generated using CRY at sea level, respectively. These distributions are taken as inputs to the G4ISMRAN simulations for studying muon-induced neutrons inside the detector setup. Figure~\ref{fig7} (a) shows the deposited energy distribution of muons passing through a PSB, fitted with a Landau distribution, convoluted with a Gaussian function to obtain the most probable value of energy deposition in a 10 cm thick PSB. In order to incorporate the detector response, a convolution of a Landau distribution with a Gaussian provides a better fit to the energy deposition spectra of cosmic muons in plastic scintillator. Cosmic muons lose energy at a rate of $\sim$ 2 MeV$\mathrm{gm^{-1}cm^{2}}$ and behave as MIPs. Consequently, the energy deposited by vertical muons in a PSB peaks at $\sim$ 20 MeV. Figure ~\ref{fig7} (b) and Fig. ~\ref{fig7} (c) illustrate the kinetic energy distribution of muon-induced secondary neutrons and proton recoil sum energy distribution ($\mathrm{E_{sum}}$) of those neutrons in the ISMRAN detectors, respectively, as muons pass through the composite sample shielding. It is observed that a significant fraction of events deposit energy in the range of 0.8 MeV to 10 MeV, which overlaps with the energy region of interest for the prompt IBD signal events. The estimated rate of muon-induced fast neutrons in Geant4, which can mimic prompt IBD signals within the energy range of 0.8 MeV to 10.0 MeV is 0.017$\%$ of muon events through proton recoil energy deposition in the detector array.


\begin{figure}[h]
\begin{center}
\includegraphics[width=16.4cm,height=5.0cm]{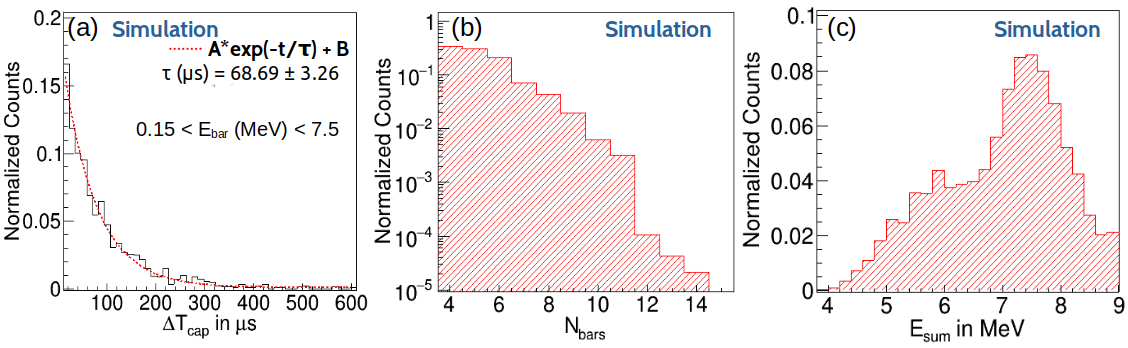}
\caption{(a) The capture time distribution of fast neutrons produced from cosmic muon interaction with the shielding structure, placed on top of the ISMRAN detector setup. The $\mathrm{N_{bars}}$ and $\mathrm{E_{sum}}$ distributions of cascades of $\gamma$-rays from these fast neutron capture events are shown in (b) and (c), respectively. All the results are from MC simulations.}
\label{fig8}
\end{center}
\end{figure} 

\subsection{Simulation of Capture Time and Energy Response of Secondary Neutrons} 
\label{sec5.3}
A few of these secondary fast neutrons get thermalized inside the PSBs and eventually get captured by Gd or H nuclei, resulting in delayed-like events inside the detector geometry. From MC simulations, Fig~\ref{fig8} (a) shows the capture time distribution of muon-induced neutron in ISMRAN, fitted with an exponential function including a flat background to obtain the mean characteristic capture time ($\tau$). The $\tau$ obtained from the fit is (68.69 $\pm$ 3.26) $\mathrm{\mu}$s, which is very similar to the capture time distribution of thermal neutrons in the IBD process~\cite{ISMRAN}. Similarly, Fig.~\ref{fig8} (b) and (c) show the number of bars hit ($\mathrm{N_{bars}}$) and $\mathrm{E_{sum}}$ distributions of the cascade of $\gamma$-rays from n-Gd capture events in the ISMRAN detector array. The percentage of excess neutron captured events in the ISMRAN geometry due to the presence of sample shielding structure is listed in table~\ref{tab1}, where, $\mathrm{N_{wos}}$ and $\mathrm{N_{ws}}$ are the number of capture events in Gd or H nuclei for without shielding and with shielding configuration, respectively. The difference in capture rates between Gd and H nuclei is due to the significantly larger thermal neutron capture cross-section of Gd nuclei (254 kbarns) compared to that of H nuclei. It is evident that with the installation of high-Z shielding material, an additional correlated background of secondary fast neutrons through proton recoil and capture events is produced inside the ISMRAN detector array~\cite{Roni}. This is due to an increase in muon-induced neutron production caused by the presence of lead material~\cite{neu_mul}. However, it is observed that the $\mathrm{\frac{N_{H}}{N_{Gd}}}$ ratio remains constant for both shielding configurations.


\begin{table}[h]
\begin{small}
\begin{center}
\caption{A comparative study of integrated yields of muon-induced neutron capture events in the detector array with and without shielding configurations, using the CRY event generator in Geant4 MC simulations for an equivalent exposure of 4.31 days in real data.}

\label{tab1}
\setlength{\arrayrulewidth}{1.0pt}
\begin{tabular}{|c|c|c|c|}
\hline
\makecell{Capture material}             & $\mathrm{N_{wos}}$ & $\mathrm{N_{ws}}$ & $\frac{(\mathrm{N_{ws}} - \mathrm{N_{wos}})}{\mathrm{N_{wos}}}$ in \%  \\ [1.5ex]
\hline
\makecell{H}                            & 294  & 484  & 64.6 \\[1.5ex]
\hline
\makecell{Gd}                           & 780  & 1290 & 65.4 \\[1.5ex]
\hline
\end{tabular}
\end{center}
\end{small}
\end{table}

\section{Measurements of Muon-induced Neutron Backgrounds in the ISMRAN Detector}
\label{6}

\subsection{Rate of Cosmic Muons in the Muon Telescope}
\label{6.1}
\begin{figure}[h]
\begin{center}
\includegraphics[width=6.6cm,height=5.8cm]{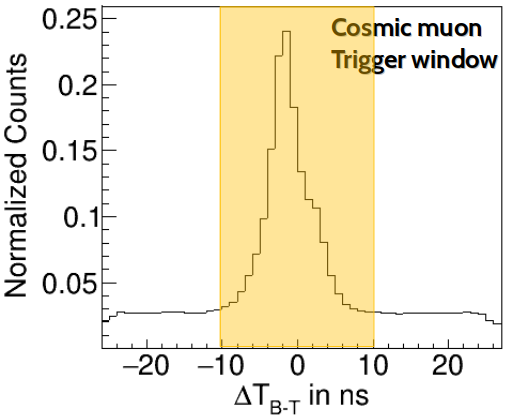}
\caption{Timestamp difference distribution ($\mathrm{\Delta T_{B-T}}$) between the two trigger paddles. The muon events are selected from the shaded region for the current analysis.}
\label{fig9}
\end{center}
\end{figure}

The muon telescope requires the passing of muons through both the top and bottom scintillator paddles, as illustrated in Fig~\ref{fig5} (b). The timestamps of the signal from both paddles are recorded independently and denoted as $\mathrm{T_{Top}}$ and $\mathrm{T_{Bottom}}$. The distribution of the timestamp difference ($\mathrm{\Delta T_{B-T}}$) between $\mathrm{T_{Top}}$ and $\mathrm{T_{Bottom}}$ is shown in Fig~\ref{fig9}. A very distinct peak between $\pm$ 10 ns in the $\mathrm{\Delta T_{B-T}}$ distribution is observed which indicates the coincidences associated with the passage of cosmic muons between these two paddles. A selection of muon-triggered events in the coincidence time window are used for the analysis to improve the purity of the cosmic muon events. The rate of cosmic muon triggered events is 1.8 $\mathrm{s^{-1}}$. The event rate of triggered cosmogenic neutrons at the paddles, derived from the evaluated efficiency-corrected ($\sim$34\%) fast neutron rate using the LS detector, is $\sim$ 0.016 $\mathrm{s^{-1}}$, which is substantially lower than the cosmic muon-induced neutron event rate of 1.8 $\mathrm{s^{-1}}$~\cite{LS_eff}. Outside this muon trigger window, where 10 $<$ $|\mathrm{\Delta T_{B-T}}|$ (ns) $<$ 20, the event rate is 0.52 $\mathrm{s^{-1}}$, which corresponds to an accidental background that may arise from random coincidences with natural $\gamma$-ray background associated with muons passing through the paddle. The rate of these accidental backgrounds, 0.52 $\mathrm{s^{-1}}$, is small compared to the muon-triggered rate of 1.8 $\mathrm{s^{-1}}$. Eventually, these accidental random backgrounds will be subtracted to determine the rate of muon-induced secondary neutron backgrounds in the ISMRAN detector. A total of $\sim$ 0.5$\times \mathrm{10^{6}}$ pure cosmic muon trigger events were identified at the trigger paddles. However, no selection cut on the integrated charge is applied on the paddles for the selection of muon events.

\begin{figure}[h]
\begin{center}
\includegraphics[width=13.2cm,height=6.2cm]{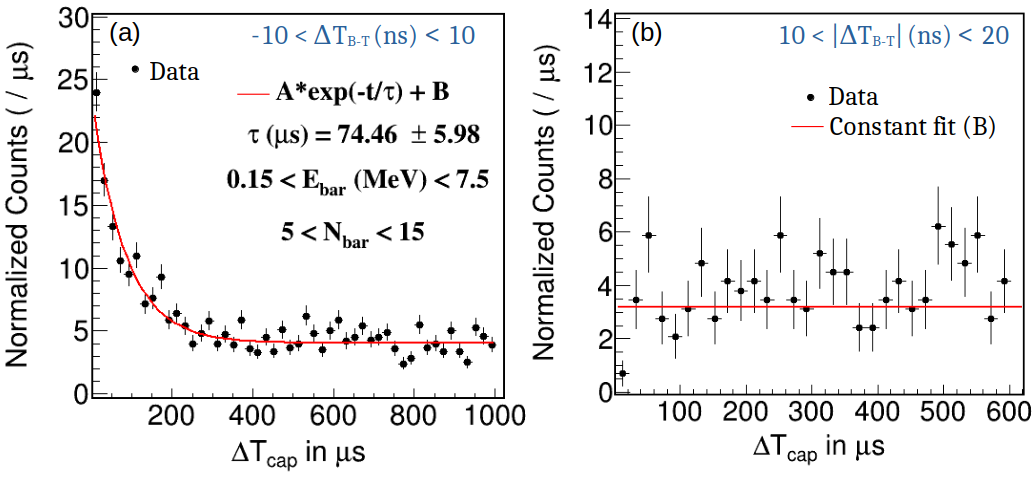}
\caption{The measured capture time distributions ($\mathrm{\Delta T_{cap}}$) between the muon trigger ($\mathrm{T_{trig}}$) event in the paddle and neutron capture event on Gd nuclei in PSBs of ISMRAN detector array, (a) inside (-10 $<$ $\mathrm{\Delta T_{B-T}}$ (ns) $<$ 10) and (b) outside (10 $<$ $|\mathrm{\Delta T_{B-T}}|$ (ns) $<$ 20) the cosmic muon trigger window, respectively. (b) Normalized per muon trigger events.}
\label{fig10}
\end{center}
\end{figure}

\subsection{Rate of Delayed n-Gd Capture Cascade $\gamma$-rays Events}
 \label{6.2}
The muon-triggered events in the plastic scintillator paddles, along with n-Gd capture events inside the ISMRAN detector array within a coincidence time window of 1000 $\mathrm{\mu}$s, are treated as delayed candidate events. Fig.~\ref{fig10} (a) and (b) show the time difference ($\mathrm{\Delta T_{cap}}$) distributions between the timestamp of muon trigger ($\mathrm{T_{Trig}}$) event in the top paddle and the timestamp of a cascade of $\gamma$-rays from n-Gd capture delayed event in the PSBs of the ISMRAN detector array, for events within the muon trigger window and for events outside the muon trigger window, respectively. Both distributions are normalized by the bin width to make the fit parameters independent of binning. The characteristic time $\tau$ is obtained by fitting the $\mathrm{\Delta T_{cap}}$ distribution in Fig.\ref{fig10} (a) with an exponential and a constant term. The exponential part represents the thermal neutron capture time on Gd nuclei and the constant term, B, shows the accidental background and from the fit we obtained the constant term of 4.37 $\pm$ 0.26 $\mathrm{\mu s^{-1}}$. The $\tau$ obtained from the fit to the neutron tagged events from the real dataset is 74.46 $\pm$ 5.98 $\mathrm{\mu}$s and is in good agreement with Geant4-based MC simulation for fast neutron events in ISMRAN. On the other hand, Fig.\ref{fig10} (b) shows a uniform distribution in $\mathrm{\Delta T_{cap}}$ with constant term (B), 3.20 $\pm$ 0.69 $\mathrm{\mu s^{-1}}$, after normalizing with muon trigger events, indicating the randomness in the muon trigger-delayed event pairs, which accounts for the purely accidental background and is also consistent with the flat component in Fig.\ref{fig10} (a), within the statistical error.



The difference in accidental background (B) rates between Fig.\ref{fig10} (a) and Fig.\ref{fig10} (b) can primarily be attributed to cosmogenic neutrons and muon-induced $\gamma$-rays backgrounds. However, the rate of these backgrounds is much lower compared to the signal muon-induced secondary neutron events. This study will be utilized to achieve the signal-to-background ratio (S/B) of these secondary neutron events in the ISMRAN experiment.


\begin{figure}[h]
\begin{center}
\includegraphics[width=12.8cm,height=6.2cm]{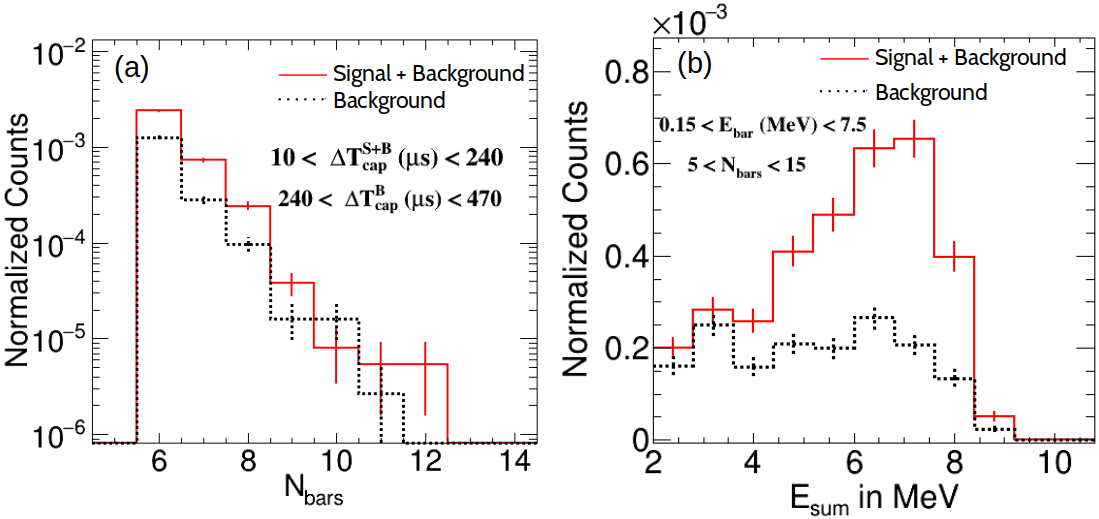}
\caption{ The comparison of measured $\mathrm{N_{bars}}$ and $\mathrm{E_{sum}}$ distributions for cascades of $\gamma$-rays from n-Gd capture events in the ISMRAN geometry are shown in (a) and (b), respectively, for the muon trigger window -10 $<$ $\mathrm{\Delta T_{B-T}}$ (ns) $<$ 10 ns, inside the signal+background and background time intervals.}
\label{fig11}
\end{center}
\end{figure} 

\begin{figure}[h]
\begin{center}
\includegraphics[width=7.8cm,height=7.2cm]{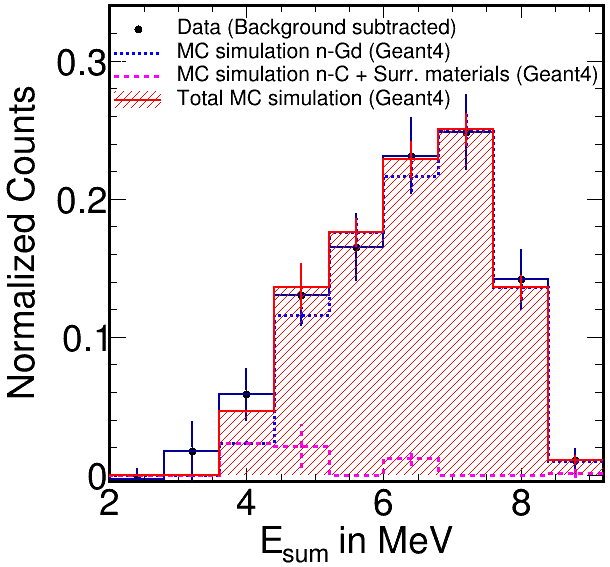}
\caption{Data-MC comparison of background subtracted $\mathrm{E_{sum}}$ distributions for n-Gd capture cascade $\gamma$-rays events, as well as neutron capture on carbon and surrounding materials, from muon-induced fast neutrons in detector array.}
\label{fig12}
\end{center}
\end{figure} 

The correlated $\mathrm{\Delta T_{cap}}$ distribution of muon trigger-delay paired events obtained from Fig.~\ref{fig10} (a) is divided into two time intervals (1) Signal + Background (S+B) interval, which ranges from 10 $<$ $\mathrm{\Delta T_{cap}}$ ($\mu$s) $<$ 240, and (2) Background (B) interval, which ranges from 240 $<$ $\mathrm{\Delta T_{cap}}$ ($\mu$s) $<$ 470. This selection is inspired by the spread of the $\mathrm{\Delta T_{cap}}$ of the neutron capture events on the Gd nuclei from Geant4 simulation results, which drops down close to zero at $\sim$240 $\mu$s. The S+B interval includes correlated muon trigger-delay paired events and accidental background events, whereas the B interval consists of mainly accidental $\gamma$ backgrounds. In this work, $\mathrm{E_{sum}}$ is defined as the reconstructed total energy of each event, obtained by summing the $\mathrm{E_{bar}}$ in each PSB~\cite{roni_ISMRAN}. Only those PSBs are selected for the $\mathrm{E_{sum}}$ distribution of delayed events where the energy $\mathrm{E_{bar}}$ in each PSB is between 0.15 MeV and 7.5 MeV. In order to reduce the contamination from the cosmic muon events, the $\mathrm{E_{sum}}$ for the n-Gd captured event is required to be in the range of 2.6 MeV to 10.0 MeV and $\mathrm{N_{bars}}$ should be in the range 6 to 14~\cite{roni_ISMRAN}. This selection criteria for $\mathrm{N_{bars}}$ increases the purity of the n-Gd capture event in the data for the secondary neutron rate estimation. From the simulations, most n-Gd capture cascade $\gamma$ events with a single high-energy $\gamma$-ray have a bar multiplicity of two to three PSBs. Our selection of the $\mathrm{N_{bars}}$ of the cascade event in the real dataset is more than 5 PSBs. The $\mathrm{N_{bars}}$ and $\mathrm{E_{sum}}$ have been compared in time intervals of (S+B) and B for the cascade of $\gamma$-rays from n-Gd capture delayed events. Figure~\ref{fig11} (a) and (b) show the measured $\mathrm{N_{bars}}$ and $\mathrm{E_{sum}}$ distributions for cascades of $\gamma$-rays from n-Gd capture events, respectively for muon-induced neutrons, compared to the accidental background events in the ISMRAN geometry. Moreover, Fig.~\ref{fig12} shows a comparison between background-subtracted data and MC simulation for the reconstructed $\mathrm{E_{sum}}$ distributions for the cascade of $\gamma$-rays from n-Gd capture events, as well as neutron capture on carbon and surrounding materials. A clear peak at 7.4 MeV is observed in the data, showing reasonable agreement with the Geant4 MC simulations within our region of interest, 6 MeV to 9 MeV. In the $\mathrm{E_{sum}}$ distribution around 4-5 MeV, a significant contribution from residual $\gamma$-ray backgrounds, possibly originating from neutron capture on carbon nuclei (n-C), has been incorporated into the simulation for comparison with the real dataset. We estimate from Geant4 that only 6$\%$ of the capture events originate from n-C interactions and neutron capture on surrounding materials. However, by applying stricter selection cuts for identifying delayed n-Gd capture events, we can effectively mitigate these background events.  



\subsection{Estimating the Reconstruction Efficiency for Muon-Induced Neutron Capture Events}
 \label{6.3}
Efficiency values were estimated based on selection criteria applied in Geant4 MC simulations, using pure n-Gd capture events. A cut-based analysis technique was employed to apply the complete set of selection criteria for the above mentioned variables to the MC event dataset in the ISMRAN geometry. Consequently, the progressive impact on event selection efficiency should be evaluated to determine the final efficiency value for these variables. A detailed estimation of efficiency is summarized in Table~\ref{tab2}, showing the various selection criteria applied to the correlated n-Gd delayed candidate events and their effects on the reconstruction efficiency.

\begin{table}[h]
\begin{small}
\begin{center}
\caption{Selection cuts and their corresponding event selection efficiencies from the Geant4-based MC simulated dataset.}
\label{tab2}
\setlength{\arrayrulewidth}{1.0pt}
\begin{tabular}{|c|c|c|}
\hline
\makecell{Variables} & {Selection criteria} &  Efficiency ($\%$) \\ [1.5ex]
\hline
\makecell{$\mathrm{\Delta T_{cap}}$} & 10 $<$ $\mathrm{\Delta T_{PD} (\mu s) }$ $<$ 240     &  85.5  \\ [1.5ex]
\hline
\makecell{$\mathrm{E_{sum}}$}        & 2.6 $<$ $\mathrm{E_{sum}}$ (MeV) $<$ 10.0                 &  65.5  \\ [1.5ex]
\hline
\makecell{$\mathrm{N_{bars}}$}       & 5 $<$ $\mathrm{N_{bars}}$ $<$ 15                    &  23.6  \\ [1.5ex]
\hline
\end{tabular}
\end{center}
\end{small}
\end{table} 

In estimating the reconstruction efficiency, we only consider events where thermal neutrons are captured by Gd nuclei. Neutron capture by H nuclei produces a 2.2 MeV monoenergetic $\gamma$-ray, which does not satisfy our $\mathrm{E_{sum}}$ selection criteria for the delayed capture event. This study describes the successive reduction in efficiency due to the various selection criteria applied to the muon-induced delayed n-Gd capture events, resulting in an overall cumulative efficiency of $\sim$ 23.6\% after implementing all the selection criteria.

\subsection{Evaluating the Rate of Muon-induced Neutrons}
\label{6.4}
The difference in the integrated yields of muon trigger-delayed event pairs in these two time intervals can be utilized for estimating the correlated cosmic muon-induced neutron events, caused by the passive sample shielding configuration of 600 $\mathrm{cm^{2}}$ surface area on top of the ISMRAN geometry. From the measured data, the obtained integrated yields of muon trigger-delayed event pairs in S+B and B regions from Fig.~\ref{fig10} (a) are $\mathrm{N_{PD}(S+B)}$ = 2837 $\pm$ 53 (stat.) and $\mathrm{N_{PD}(B)}$ = 1457 $\pm$ 38 (stat.), respectively. The errors quoted here are only statistical in nature.

For the estimation of systematic uncertainties associated with secondary neutrons produced by cosmic muons, we have included the following sources of uncertainty. These consist of uncertainties in the reconstruction of $\mathrm{E_{sum}}$ for n-Gd capture delayed events using individual $\mathrm{E_{bar}}$ with different thresholds in the data, cosmic muon trigger efficiency using the paddle trigger, the estimation of cosmic muon flux, coincidence timing selection between trigger muons and delayed n-Gd capture events, and uncertainties arising from Data-MC discrepancies in the n-Gd capture model using Geant4 simulations.

The delayed $\mathrm{E_{sum}}$ reconstruction is affected by the individual $\mathrm{E_{bar}}$ selection thresholds, which vary from 0.15 MeV to 0.3 MeV, leading to a $\sim$3$\%$ variation across the 90 PSBs in the ISMRAN detector array~\cite{roni_ISMRAN}. This uncertainty arises due to varying thresholds in individual PSBs, implemented to remove low-energy pedestal events. The individual thresholds have been propagated in Geant4 simulations to estimate the uncertainty due to threshold variations. From the Geant4 simulations shown in Fig.\ref{fig8} (a), it is observed that $\mathrm{\Delta T_{cap}}$ decreases exponentially, with the contribution to the neutron capture events becoming negligible above $\sim$240 $\mu$s. This exponential decay introduces uncertainty in selecting the signal and background regions, when the region of interest is defined for $\mathrm{\Delta T_{cap}}$ $>$ 240 $\mu$s. To estimate this uncertainty, we define two regions for $\mathrm{\Delta T_{cap}}$: 10 $<$ $\mathrm{\Delta T_{cap}}$ ($\mu$s) $<$ 200 and 10 $<$ $\mathrm{\Delta T_{cap}}$ ($\mu$s) $<$ 280. The differences in the yields obtained from these two regions are used to quantify the systematic uncertainty associated with the $\mathrm{\Delta T_{cap}}$ selection for identifying correlated delayed n-Gd capture events. The lower limit of $\mathrm{\Delta T_{cap}}$ is set at 200 $\mu$s because, as illustrated in Fig.~\ref{fig10}(a), the signal-to-background ratios are nearly equal at this threshold. This equality may introduce uncertainty in the secondary neutron yield during the selection process for $\mathrm{\Delta T_{cap}}$ variable. It introduces an additional systematic uncertainity of 2 $\%$ in selection of signal and background events. Because of the inadequate modeling of the n-Gd capture cascade $\gamma$-rays in Geant4, we used DICEBOX as a standalone tool to generate $\gamma$-ray cascades. A conservative 2$\%$ systematic uncertainty is assigned to the n-Gd $\gamma$-ray cascade production, which, in turn, affects the reconstruction and efficiency calculations~\cite{neu_mul}.


Table~\ref{tab3} summarizes the systematic uncertainties associated with the selection criteria, which are propagated to the final muon-induced secondary neutron yield. A total systematic uncertainty of 5.3\% is estimated for the 4.31 days of data recorded in the ISMRAN geometry.

\begin{equation}\label{event1}
\mathrm{N^{cor}_{n-Gd} = N_{PD} (S+B) - N_{PD} (B)}  
\end{equation}




\begin{table}[h]
\begin{small}
\begin{center}
\caption{Systematic errors in the experimental data and Geant4 simulations, propagated to muon-induced secondary neutron capture events on Gd nuclei.}
\label{tab3}
\setlength{\arrayrulewidth}{1.0pt}
\begin{tabular}{|c|c|}
\hline
\makecell{Error Source} & Systematic Error in $\%$  \\ 
\hline
\makecell{$\mathrm{E_{bar}}$}                       & 3                          \\[1.0ex]
\hline
\makecell{$\mathrm{\Delta T_{cap}}$}                & 2                          \\[1.0ex]
\hline
\makecell{Cosmic Muon Trigger Efficency}            & 1.3                        \\[1.0ex]
\hline
\makecell{Cosmic Muon Flux}                         & 3                          \\[1.0ex]
\hline
\makecell{DICEBOX model}                            & 2                         \\[1.0ex]
\hline
\makecell{\bf{Total Systematic}}                    & \bf{5.3}                         \\[1.0ex]
\hline
\end{tabular}
\end{center}
\end{small}
\end{table}


The yield of correlated n-Gd capture events ($\mathrm{N^{cor}_{n-Gd}}$) for 4.31 days of collected dataset, obtained from the above mentioned statistical background subtraction method is $\mathrm{N^{cor}_{n-Gd}}$ (S) = 1380 $\pm$ 66 (stat.), ~\cite{miniISMRAN}. Our results are compared to the theoretical estimation obtained from Geant4-based MC simulation (similar size to the amount of data taken for 4.31 days). The expected correlated cosmic muon-induced neutron events in the ISMRAN geometry is 1290 $\pm$ 36 (stat.) for the sample shielding structure of the above mentioned surface area (600 $\mathrm{cm^{2}}$) in Geant4, and is in reasonable agreement with the measured data within the errors. The presence of fast neutron events involving high-energy cosmic muons is clearly observed with a S/B ratio of $\sim$0.95. The final yield of correlated n-Gd capture events is determined by applying correction for selection cut efficiency of 24\%, which introduces a source of systematic uncertainty. In this study, we report that the measured, efficiency-corrected cosmic muon-induced neutron yield in the ISMRAN geometry per day is 1334 $\pm$ 64 (stat.) $\pm$ 70 (sys.).

In order to estimate the total fast neutron yield for the actual shielding structure of a surface area 9000 $\mathrm{cm^{2}}$ (90 $\times$ 100 $\mathrm{cm^{2}}$), the acceptance correction has been computed using Geant4 simulation. Muon-induced neutron yields have been estimated for both actual shielding and the sample shielding structure. Subsequently, normalizing the surface area (9000/600), an acceptance correction factor can be derived, and it comes out to be 0.4 in our study. Eventually, muon-induced neutron yields from real data for the sample shielding structure have been corrected with this model-dependent acceptance factor. These experimental results and the acceptance correction factor enable us to assess the total expected number of cosmic muon-induced neutron capture events in the ISMRAN geometry, which is 3335 $\pm$ 160 (stat.) $\pm$ 175 (sys.) neutrons $\mathrm{day^{-1}}$ for a top shielding configuration of surface area 9000 $\mathrm{cm^{2}}$. These estimated numbers are also consistent with the measured correlated background yields during reactor OFF (ROFF) condition with the ISMRAN detector array inside the reactor facility~\cite{prospect_bkg}.

Finally, the neutron production yield ($\mathrm{Y_{n}}$) from cosmic muon in the ISMRAN geometry has been evaluated using the measured results from the current study, representing the average number of neutrons generated per muon interaction with composite sample shielding. This yield can be determined if the average distance ($\mathrm{L_{avg}}$) traversed by muons and the average interaction density ($\mathrm{\rho_{avg}}$) of muons through the shielding materials are known. We are investigating the spallation neutron products induced by cosmic muons with an average energy of $\sim$4 GeV, as they pass through the sample shielding, assuming an angular distribution of $\mathrm{cos^{2}(\theta)}$. 

The total average path length ( $\mathrm{ L_{avg} }$ = [$\mathrm{L_{avg}}]_{\mathrm{Pb}}$ + [$\mathrm{L_{avg}}]_{\mathrm{BPE}}$ ) of cosmic muons in the Pb + BPE target is estimated from Geant4 simulation to be 25.6 cm, with an average interaction target density $\left( \rho_{\mathrm{avg}} = \frac{[\rho_{\mathrm{avg}}]_{\mathrm{Pb}} \times [L_{\mathrm{avg}}]_{\mathrm{Pb}} + [\rho_{\mathrm{avg}}]_{\mathrm{BPE}} \times [L_{\mathrm{avg}}]_{\mathrm{BPE}}}{L_{\mathrm{avg}}} \right)$ of 6.3 g/$\mathrm{cm^{3}}$ ~\cite{rate_neutron}. The muon-induced neutron production yield is then defined as,




\begin{equation}\label{event5}
\mathrm{Y_{n} = \left (\frac{1}{L_{avg}\times\rho_{avg}} \right ). \left (\frac{\mathrm{N_{ncap}}}{\eta}\right ). \left (\frac{1}{\mathrm{N_{\mu}}}\right ) } .
\end{equation}

Here, $\mathrm{N_{ncap}}$ represents the measured efficiency-corrected number of neutron events ($\sim$ 1334) per day, obtained in this study, resulting from cosmic muon interactions with combined sample shielding and subsequently captured in the ISMRAN geometry after thermalization in PSBs. The $\eta$ represents the ratio of efficiency-corrected neutron capture events to the total muon-generated neutron events, estimated to be 34\% from Geant4 simulation. The $\mathrm{N_{\mu}}$ corresponds to the number of cosmic muons passing through the sample shielding with a surface area of 600 $\mathrm{cm^{2}}$ per day, which is 864000. Therefore, the production yield of secondary neutrons in the ISMRAN geometry due to the sample shielding is 2.81$\times$$\mathrm{10^{-5}}$ neutrons/$\mu$/(g/$\mathrm{cm^{2}}$). 

 
\section{Conclusions and Outlook}
\label{sec7}
One of the most significant backgrounds for any reactor-based, above-ground ${\overline{\ensuremath{\nu}}}_{e}$ experiment is generated by fast neutrons. Due to their indistinguishable signatures from those obtained by an inverse beta decay process of ${\overline{\ensuremath{\nu}}}_{e}$, it is essential to precisely estimate these backgrounds. In an above-ground, very short baseline reactor ${\overline{\ensuremath{\nu}}}_{e}$ experiment, the active detectors need to be shielded to substantially suppress the reactor-induced $\gamma$-rays and fast neutron backgrounds. These shielding materials, typically made of lead (Pb) and boronated polyethylene (BPE), can generate a secondary fast neutron through the interaction with high-energy cosmic muons, which can easily penetrate the shielding materials and leave signature in terms of deposited energy in the active detector. In the current study, a muon-triggering method is presented to evaluate the yields of these correlated secondary neutron backgrounds with a large area segmented plastic scintillator detector array. These secondary fast neutrons are mainly produced via the muon-nucleus spallation interaction of high-energy cosmic muon with 10 cm thick Pb and 10 cm thick BPE shielding structure. This work focuses on the detection mechanism of secondary neutrons produced by cosmic muons, which are detected through cascades of $\gamma$-rays emitted by neutron capture on the Gd sheet in PSBs. The measured deposited energy and capture time responses of these secondary neutrons in PSBs of the ISMRAN has been compared with the Geant4-based MC simulations. The measured mean characteristic capture time ($\tau$) is 74.46 $\pm$ 5.98 $\mathrm{\mu}$s and which is consistent with the MC simulation results. In addition, a reasonable agreement between data and MC simulated events is observed for the $\mathrm{E_{sum}}$ distribution of correlated n-Gd capture cascade $\gamma$-ray events. The statistical subtraction method is employed to obtain the measured fast neutron yield of 1380 $\pm$ 66 (stat.), produced by cosmic muons. Reconstruction efficiency has been estimated at 24\% using Geant4 simulation, to obtain the actual (efficiency-corrected) secondary neutron yield and evaluate the systematic uncertainty associated with the selection criteria. Due to the presence of a passive high-density sample shielding structure with a surface area of 600 $\mathrm{cm^{2}}$, installed on top of the ISMRAN detector array, resulted in 1334 $\pm$ 64 (stat.) $\pm$ 70 (sys.) fast neutron capture events per day, which is also consistent with the MC simulation results obtained from Geant4, within the errors. From these studies, we estimated that the rate of cosmic muon-induced neutron capture events in the ISMRAN geometry for the actual shielding configuration of 9000 $\mathrm{cm^{2}}$ surface area is 3335 $\pm$ 160 (stat.) $\pm$ 175 (sys.) neutrons $\mathrm{day^{-1}}$, using the extrapolation method. These experimental results also allow us to select a cosmic muon rejection time window of 240 $\mathrm{\mu}$s after the muon passage for vetoing muon-induced background events. Finally, the calculated neutron production yield in the ISMRAN geometry, for a configuration of 10 cm thick Pb followed by 10 cm thick BPE is 2.81$\times$$\mathrm{10^{-5}}$ neutrons/$\mu$/(g/$\mathrm{cm^{2}}$). The expected rate of muon-induced neutron background can be mitigated by using the segmentation of the ISMRAN geometry, along with the implementation of advanced machine learning algorithms~\cite{MLP}, to reduce experimental uncertainties and enhance the physics sensitivity for detecting true ${\overline{\ensuremath{\nu}}}_{e}$ events in the ISMRAN detector array inside the Dhruva reactor facility.


\section{Acknowledgments}
We are grateful to the EmA$\&$ID workshop in BARC for building the temporary structure for placing the ISMRAN detectors in the detector laboratory. We would also like to thank Dr. U. K. Pal, and Dr. P. C. Rout, NPD BARC for providing us the $\mathrm{CeBr_{3}}$ and NE213 liquid scintillator detectors for the background measurements. 


\end{document}